\documentclass[12pt,aps,nofootinbib]{revtex4}
\usepackage{amsmath}
\usepackage{graphicx}

\newcommand{\Gammamat}{\boldsymbol \Gamma}
\newcommand{\Rmat}{\mathbf R}
\renewcommand{\vec}[1]{\boldsymbol #1}
\DeclareMathOperator{\trace}{tr}
\DeclareMathOperator{\Trace}{Tr}

\newcommand{\qT}{q_{\scriptscriptstyle T}}

\begin{document}

\title{Functional renormalisation group for two-body scattering}
\author{Michael C. Birse}

\affiliation{Theoretical Physics Group, School of Physics and Astronomy,
The University of Manchester, M13 9PL, UK}
\date{\today}
\begin{abstract}
The functional renormalisation group is applied to the effective action
for scattering of two nonrelativistic fermions. The resulting physical 
effective action is shown to contain the correct threshold singularity.
The corresponding ``bare" action respects Galilean invariance up to 
second order in momenta. Beyond that order it contains terms that violate
this symmetry and, for the particular regulator considered, nonanalytic 
third-order terms. The corresponding potential can be expanded around a 
nontrivial fixed point using the power counting appropriate to a system 
with large scattering length.
\end{abstract}
\maketitle

The functional or ``exact" renormalisation group (RG) is proving to 
be a powerful tool for the study of phase transitions in a variety of 
systems, from condensed matter to particle physics \cite{BTW02,DMT03}.
Recently there has been particular interest in applying it to pairing 
in dense matter consisting of nonrelativistic fermions, such as nuclear 
matter or cold atoms in traps \cite{BKMW05,kri06,DGPW07}. In this context, 
it has been used to examine the cross-over between BCS superfluidity and 
Bose-Einstein condensation of tightly bound pairs. 

In this approach, a scale-dependent regulator is used to interpolate 
between a bare action and the physical effective action. This regulator 
is designed to suppress the contributions of fluctations with momenta 
smaller than the cut-off scale, $k$. As $k\rightarrow 0$, it vanishes
so that all fluctuations are integrated out and we arrive at the full 
effective action describing physical scattering. The starting point for 
the RG flow is a suitably parametrised ``bare" action at some large starting 
scale $K$. This is often expressed as an expansion in derivatives of the 
fields. To fix the parameters in it, we can evolve down to the physical 
limit in vacuum and there we can match, for example, to the two-body 
scattering matrix.

The input into these calculations of fermionic matter is thus in the form 
of two-body (and ultimately also three- and more-body) forces between the
particles in vacuum. This means that there is a close connection to the 
effective field theories that are being developed for few-body systems, 
especially in nuclear physics \cite{BvK02,epel06} and, in particular, to 
the Wilsonian RG methods that have been developed to determine the power 
countings in these theories \cite{BMR99}. In fact, the Kyushu group has 
applied the functional RG to nucleon-nucleon scattering \cite{HKN07} and 
shown that the results agree with the Wilsonian approach. A first application 
to three-body systems has been made more recently \cite{DKS07}.

Here I apply the ``exact" RG to the scattering of two nonrelativistic 
fermions, using the version based on the Legendre-transformed effective 
action \cite{BTW02,DMT03} and introducing a boson field to describe 
correlated pairs of fermions (``dimers") as in Refs.~\cite{BKMW05,kri06,DGPW07}.
In the present work I do not make a derivative expansion of the boson 
self-energy but instead I keep terms to all orders in energy and momenta. 
This provides an extension of the calculations in Ref.~\cite{HKN07}
and Sec.~IV of \cite{DKS07} that allows both an analysis of the 
scaling properties, as in Ref.~\cite{BMR99}, and the identification of 
artefacts caused by the regulator, such as violations of Galilean 
invariance.

In describing systems of fermions with attractive forces, it is
convenient to represent the interactions in terms of boson
fields. These fields describe correlated, but not necessarily bound, 
pairs of fermions, like the dibaryon fields often used in nuclear 
effective theories \cite{kap97}. Here I consider a single type of 
fermion, with two spin states, and I introduce one boson field to 
describe the interactions between them. The Fourier transforms of 
these fields are denoted by $\widetilde\psi(q)$ and $\widetilde\phi(q)$, 
respectively. The starting point is the following ansatz for the 
effective action $\Gamma$,
\begin{eqnarray}
\Gamma[\psi,\psi^\dagger,\phi,\phi^\dagger;k]&=&\int {\rm d}^4\!q\,\left[
\widetilde\phi(q)^\dagger\,\Pi(q_0,\vec q;k)\,
\widetilde\phi(q)+\widetilde\psi(q)^\dagger\left(q_0-\frac{\vec q^2}{2M}\right)
\widetilde\psi(q)\right]\cr
\noalign{\vskip 5pt}
&&
- g\,\frac{1}{(2\pi)^2}\,\int {\rm d}^4\!q_1\,{\rm d}^4\!q_2\,
\left(\frac{\mbox{i}}{2}\,\widetilde\phi(q_1+q_2)^\dagger
\widetilde\psi(q_2)^{\rm T}\sigma_2
\widetilde\psi(q_1)\right.\cr
\noalign{\vskip 5pt}
&&\qquad\qquad\qquad\qquad\left.-\frac{\mbox{i}}{2}\,
\widetilde\psi(q_1)^\dagger\sigma_2
\widetilde\psi(q_2)^{\dagger{\rm T}}\widetilde\phi(q_1+q_2)\right)+\cdots.
\label{eq:ansatz}
\end{eqnarray}
This is essentially the same as the action considered in 
Refs.~\cite{BKMW05,kri06,DGPW07,DKS07}, except for the inverse boson 
propagator $\Pi(q_0,\vec q;k)$ which I do not expand in powers of energy or 
momentum. 

In Eq.~(\ref{eq:ansatz}), I have chosen to express the interactions 
between the fermions entirely in terms of the boson field. The action 
thus does not include any explicit two-body interactions and one can 
regard $\phi$ as an auxiliary field arising from bosonisation
of these terms. If I make this choice at the starting scale $K$ then, 
at least in vacuum, it is maintained during the RG evolution; there 
are no one-particle-irreducible loop diagrams that generate two-body 
interactions between the fermions. In vacuum, there are also other 
simplifications. In particular, conservation of fermion number means 
that there are no boson loops that dress either the fermion propagator 
or the fermion-boson coupling and so these retain their free forms.

The action evolves in the usual way as the regulator scale is changed 
\cite{BTW02,DMT03}:
\begin{equation}
\partial_k\Gamma=+\frac{\mbox{i}}{2}\,\Trace \left[(\partial_k\Rmat_F)\,
\left((\Gammamat ^{(2)}-\Rmat)^{-1}\right)_{FF}\right]
-\frac{\mbox{i}}{2}\,\Trace \left[(\partial_k\Rmat_B)\,
\left((\Gammamat ^{(2)}-\Rmat)^{-1}\right)_{BB}\right],
\label{eq:evolution}
\end{equation}
where $\Gammamat ^{(2)}$ is the matrix of second derivatives of the action with 
respect to the fields. In vacuum we shall need only the fermionic part, which 
involves 
\begin{equation}
\left(\Gammamat ^{(2)}-\Rmat\right)_{FF}
=\left(\begin{array}{cc}
q_0-\displaystyle{\frac{\vec q^2}{2 M}}-R_F(\vec q,k)&ig\phi\sigma_2\cr
\noalign{\vskip 5pt}
-ig\phi^\dagger\sigma_2&q_0+\displaystyle{\frac{\vec q^2}{2 M}}
+R_F(\vec q,k)\end{array}\right).
\label{eq:g2ff}
\end{equation}
Note that, although no approximation has yet been made, the driving terms 
on the right-hand side of Eq.~(\ref{eq:evolution}) have the form of 
one-loop integrals. 

The evolution of the boson self-energy can be found from
\begin{equation}
\partial_k\Pi(P_0,\vec P;k)=\left.\frac{\delta^2}{\delta\widetilde\phi(P)\,
\delta\widetilde\phi(P)^\dagger}\,\partial_k\Gamma\right|_{\tilde\phi=0},
\end{equation}
where, in vacuum, only fermion loops contribute:
\begin{eqnarray}
\left.\frac{\delta^2}{\delta\widetilde\phi(P)\,
\delta\widetilde\phi(P)^\dagger}\,\partial_k\Gamma
\right|_{\widetilde\phi=0}&=&+\mbox{i}\,\Trace \left[(\partial_k\Rmat_F)\,
(\Gammamat ^{(2)}_{FF}-\Rmat_F)^{-1}
\,\Gammamat ^{(3)\dagger}_{FF\phi}\,
(\Gammamat ^{(2)}_{FF}-\Rmat_F)^{-1}\right.\cr
&&\qquad\qquad\qquad\qquad\left.\times\,\Gammamat ^{(3)}_{FF\phi}\,
(\Gammamat ^{(2)}_{FF}-\Rmat_F)^{-1}
\right].
\end{eqnarray}
Here the necessary third derivatives of $\Gamma$ are
\begin{equation}
\Gammamat ^{(3)}_{FF\phi}
=\frac{\partial}{\partial\widetilde\phi}\,\Gammamat ^{(2)}_{FF}
=\frac{1}{(2\pi)^2}\left(\begin{array}{cc} 0&\mbox{i}g\sigma_2\cr
0&0\end{array}\right),
\end{equation}
and the matrix trace gives
\begin{eqnarray}
&&\trace\left[(\partial_k\Rmat_F)\,
(\Gammamat ^{(2)}_{FF}-\Rmat_F)^{-1}
\,\Gammamat ^{(3)\dagger}_{FF\phi}\,
(\Gammamat ^{(2)}_{FF}-\Rmat_F)^{-1}
\,\Gammamat ^{(3)}_{FF\phi}\,
(\Gammamat ^{(2)}_{FF}-\Rmat_F)^{-1}\right]\cr
\noalign{\vskip 5pt}
&&\quad=-2g^2\,\frac{1}{(2\pi)^4}\,\frac{\partial_kR_F(\vec q,k)}
{\Bigl[q_0+E_{FR}(\vec q,k)-\mbox{i}\epsilon\Bigr]^2
\Bigl[q_0+P_0-E_{FR}(\vec q+\vec P,k)+\mbox{i}\epsilon\Bigr]},
\end{eqnarray}
where
\begin{equation}
E_{FR}(\vec q,k)=\vec q^2/2M+R_F(\vec q,k).
\end{equation}

After performing the contour integral over $q_0$ and symmetrically
routing the external momentum $\vec P$ (by shifting $\vec q\rightarrow 
\vec q-\vec P/2$ and using the symmetry under $\vec P\rightarrow -\vec P$), 
the driving term can be written in the 
form of a derivative with respect to $k$:
\begin{equation}
\left.\frac{\delta^2}{\delta\widetilde\phi(P)\,
\delta\widetilde\phi(P)^\dagger}\,\partial_k\Gamma
\right|_{\eta=0}=g^2\partial_k\int\frac{{\rm d}^3\!{\vec q}}{(2\pi)^3}\,
\frac{1}{E_{FR}(\vec q-\vec P/2,k)+E_{FR}(\vec q+\vec P/2,k)-P_0
-\mbox{i}\epsilon}.
\end{equation}
This allows us to relate $\Pi(P_0,P;k)$ to the physical self-energy 
(at $k=0$), by integrating the evolution equation to get
\begin{eqnarray}
\Pi(P_0,P;k)&-&\Pi(P_0,P;0)\cr
\noalign{\vskip 5pt}
&=&-g^2\int\frac{{\rm d}^3\!{\vec q}}{(2\pi)^3}\,\left\{
\frac{1}{E_{FR}(\vec q-\vec P/2,0)+E_{FR}(\vec q+\vec P/2,0)-P_0
-\mbox{i}\epsilon}\right.\cr
\noalign{\vskip 5pt}
&&\qquad\qquad\qquad\quad\left.
-\,\frac{1}{E_{FR}(\vec q-\vec P/2,k)+E_{FR}(\vec q+\vec P/2,k)-P_0
-\mbox{i}\epsilon}\right\}.\cr
&&\label{eq:Piphys}
\end{eqnarray}
(From now on, all expressions are three-dimensional and $P$ denotes the 
magnitude of $\vec P$.)

A convenient choice of regulator is the one used by Diehl 
\textit{et al.} \cite{DGPW07},
\begin{equation}
R_F(\vec q,k)=\frac{k^2-q^2}{2M}\,\theta(k-q).
\label{eq:RF}
\end{equation} 
This is just the nonrelativistic version of the ``optimised" cut-off for
boson fields introduced by Litim \cite{lit01}.
It is a variant of the sharp cut-off which has the advantage of leading 
to a momentum-independent fermion propagator for three-momenta smaller 
than $k$. With it, the integrand in Eq.~(\ref{eq:Piphys}) vanishes outside 
the intersecting spheres, $|\vec q\pm \vec P/2|\leq k$. In the region 
between the spheres, the integrand for $q_z\geq 0$ is
\begin{equation}
\frac{M}{\qT^2+q_z^2+P^2/4-MP_0}-\frac{2M}{\qT^2+(q_z+P/2)^2+k^2-2MP_0},
\end{equation}
where I have chosen the $z$-axis along the direction of $\vec P$ and I have 
defined $\qT=\sqrt{q_x^2+q_y^2}$. It is convenient to integrate this over the 
union of the two spheres, $0\leq q_z\leq k+P/2$, $0\leq \qT\leq 
\sqrt{k^2-(q_z-P/2)^2}$. The result can be corrected by adding the 
integral of
\begin{equation}
\frac{2M}{\qT^2+(q_z+P/2)^2+k^2-2MP_0}-\frac{M}{k^2-MP_0},
\end{equation}
over the intersection of the spheres, $0\leq q_z\leq k-P/2$, 
$0\leq \qT\leq \sqrt{k^2-(q_z+P/2)^2}$. The symmetry of the integrand
means that the regions with $q_z\leq 0$ make identical contributions to the 
corresponding ones with $q_z\geq 0$.

After integrating over $\vec q$, the resulting form of the self-energy is
\begin{eqnarray}
\Pi(P_0,P;k)&-&\Pi(P_0,P;0)\cr
\noalign{\vskip 5pt}
&=&-\,\frac{g^2M}{4\pi^2}\Biggl\{\mbox{i}\pi\sqrt{MP_0-P^2/4}\cr
\noalign{\vskip 5pt}
&&\qquad\qquad-\,\sqrt{MP_0-P^2/4}\,
\ln\left(\frac{k+P/2+\sqrt{MP_0-P^2/4}}{k+P/2-\sqrt{MP_0-P^2/4}}\right)\cr
\noalign{\vskip 5pt}
&&\qquad\qquad+\,\frac{1}{k^2-MP_0}\left[
\frac{10}{3}\,k^3-4kMP_0-\frac{3}{2}\,k^2P+2MP_0P-\frac{P^3}{24}\right]\cr
\noalign{\vskip 5pt}
&&\qquad\qquad+\,4\sqrt{k^2-2MP_0}\left[\arctan\left(
\frac{k+P}{\sqrt{k^2-2MP_0}}\right)\right.\cr
\noalign{\vskip 5pt}
&&\mbox{\hspace{5 cm}}\left.-\arctan\left(
\frac{k}{\sqrt{k^2-2MP_0}}\right)\right]\cr
\noalign{\vskip 5pt}
&&\qquad\qquad-\,\frac{k^2-P^2-MP_0}{P}\ln\left[
\frac{k^2+kP+P^2/2-MP_0}{k^2-MP_0}\right]\Biggr\}.
\label{eq:Piphysint}
\end{eqnarray}
Here I have assumed that the scale $k$ is large ($k>\sqrt{2MP_0}$).

The on-shell fermion-fermion scattering amplitude is directly related
to the boson self energy in the physical limit ($k\rightarrow 0$) by 
\begin{equation}
\frac{1}{T(p)}=\frac{1}{g^2}\,\Pi(P_0,P;0),
\end{equation}
where $p=\sqrt{MP_0-P^2/4}$ is the relative momentum of the two fermions.
The effective-range expansion \cite{newton} can be used to write this 
amplitude as
\begin{equation}
\frac{1}{T(p)}=-\,\frac{M}{4\pi}\left(-\mbox{i}p-\frac{1}{a}
+\frac{1}{2}r_e p^2+\cdots\right).
\end{equation}
Comparing this with the result (\ref{eq:Piphysint}), we see that 
the RG evolution generates the correct nonanalytic term 
describing the threshold ($\propto\mbox{i}\sqrt{MP_0-P^2/4}$). As a result 
this cancels to leave a self-energy $\Pi(P_0,P;k)$ that is real and can be 
expanded as a power series in energy and momentum for large $k$.

Inserting the physical amplitude into Eq.~(\ref{eq:Piphysint}) and 
expanding in powers of $P_0$ and $P$, we find that $\Pi(P_0,P;k)$
can be written
\begin{equation}
\Pi(P_0,P;k)=\frac{g^2M}{4\pi^2}\,\Biggl\{-\,\frac{4}{3}\,k+\frac{\pi}{a}
+\left(\frac{8}{3k}-\frac{\pi}{2}\,r_e\right)\Bigl(MP_0-P^2/4\Bigr)
-\frac{1}{24k^2}\,P^3+\cdots\Biggr\}.
\end{equation}
The use of a regulator that does not respect Galilean invariance means 
that the evolution generates terms that are not invariant. To cancel these 
and leave the correct physical amplitude at $k=0$, the self-energy
at the starting scale $k=K$ must also contain such terms. The leading 
term that violates Galilean invariance is of order $P^3$, and others 
appear at all higher orders. 

The order-$P^3$ term is also nonanalytic in $P^2$ and so could not be 
generated by a power of $\nabla^2$. Such terms are artefacts of 
regulators like the sharp cut-off used here, Eq.~(\ref{eq:RF}). As 
discussed in Refs.~\cite{mor94,lit05,mor05}, they arise from the 
nonlocalities introduced by the discontinuity at the surface $q=k$. 
The cut-offs used here and in Refs.~\cite{lit01,DGPW07} vanish
at this surface and so the first such term appears at order $P^3$
\cite{mor05}. A cut-off with a stonger discontinuity, for example
\begin{equation}
R_F(\vec q,k)=\frac{k^2}{2M}\,\theta(k-q),
\end{equation} 
would lead to nonanalytic dependence appearing at order $P$ 
\cite{mor94}.

As in Ref.~\cite{BKMW05}, we can define a boson wave-function 
renormalisation factor by
\begin{equation}
Z_\phi(k)=\left.\frac{\partial}{\partial P_0}\,\Pi(P_0,P;k)
\right|_{P_0=P=0},
\end{equation}
and a kinetic-mass renormalisation by
\begin{equation}
\frac{1}{4M}\,Z_m(k)=-\,\left.\frac{\partial}{\partial P^2}\,\Pi(P_0,P;k)
\right|_{P_0=P=0}.
\end{equation}
From the expansion of $\Pi(P_0,\vec P;k)$ above, we find that these are
\begin{equation}
Z_\phi(k)=Z_m(k)=\frac{g^2M^2}{4\pi^2}
\left(\frac{8}{3k}-\frac{\pi}{2}\,r_e\right).
\label{eq:ZphiZm}
\end{equation}
The fact that $Z_\phi(k)$ and $Z_m(k)$ are identical is a consquence of 
the Galilean invariance of the vacuum (which is respected at this order). 
This will not be the case for calculations of fermionic matter as in 
Refs.~\cite{BKMW05,kri06,DGPW07}, where there is a preferred rest frame. 
These factors diverge as $k\rightarrow 0$ as a result of the nonanalytic 
energy-dependence of the threshold term in the physical amplitude. This 
provides a warning that a derivative expansion may not be valid for the 
physical effective action, even it can be used to parametrise the input 
bare action.

Evaluated at the starting scale $K$, the expressions (\ref{eq:ZphiZm})
can be used to provide initial conditions on the RG evolution. They show 
that the physical effective range is encoded in the initial values for 
these renormalisation factors. Note that their initial values can become 
negative if the effective range $r_e$ is positive and we try to start from 
a scale $K>16/3\pi r_e$ (i.e.~above the scale of the underlying physics).

For a starting scale $K$ larger than the momenta of interest, we can 
relate $\Pi(P_0,P;K)$ to an effective potential as in Ref.~\cite{HKN07}. 
Allowing for violations of Galilean invariance, this potential is
defined by
\begin{eqnarray}
\frac{1}{V(p,P;K)}&=&\frac{1}{g^2}\,\Pi\left((p^2+P^2/4)/M,P;K\right)\cr
\noalign{\vskip 5pt}
&=&\frac{M}{4\pi^2}\,\Biggl\{-\,\frac{4}{3}\,K+\frac{\pi}{a}
+\left(\frac{8}{3K}-\frac{\pi}{2}\,r_e\right)p^2
-\frac{1}{24K^2}\,P^3+\cdots\Biggr\}.
\end{eqnarray}
This can then be used, following Ref.~\cite{BMR99}, to define a rescaled 
potential $\hat V=(MK/2\pi^2)V$ and dimensionless momentum variables
$\hat p=p/K$, and $\hat P=P/K$. From this we see that the part of $\hat V$ 
coming from the integral in Eq.~(\ref{eq:Piphysint}) is independent of
$K$. As already noted in the conext of a derivative expansion \cite{HKN07}
of the action, this is the nontrivial fixed point of the RG, corresponding 
to a system with infinite scattering length. The different regulator means 
that detailed form of the fixed point differs from that found with a simple 
sharp cut-off \cite{BMR99}, but the fact that it contains a constant term of 
order unity (in this case $-2/3$) is universal. Similarly, the terms in
the potential from the effective-range expansion scale with the same 
powers of $K$ as found in Ref.~\cite{BMR99}. The resulting power counting
is the usual one for systems with large scattering lengths \cite{BvK02}.

This calculation of the untruncated two-body effective action illustrates 
several features of more general significance for applications of the
functional RG to few-body systems and fermionic matter. The lack of Galilean invariance of the regulator means that non-invariant terms must be included
in the initial bare action. With a sharp cut-off like the one used here, 
the bare actions will also need to contain terms with nonanalytic dependences 
on $P^2$, which would not be present in a standard derivative expansion.
The need for such unphysical terms, with carefully tuned coefficents, is a
potentially significant complication for applications of this approach
to matter. It has not had to be addressed so far, but only because the 
actions used have been truncated at too low orders for these effects to 
arise \cite{BKMW05,DGPW07}.

The RG evolution generates the correct square-root singularity associated 
with the threshold for two-body scattering. However, although this is easy 
to identify in the analytic solution of the RG equations presented here, 
the corresponding singularities will require careful treatment as the 
physical limit is approached in numerical integrations. For example, if 
the action is expanded in powers of derivatives, the resulting boson 
wave-function renormalisation diverges as $k\rightarrow 0$.

Finally, the two-body action for large scales can be expressed in terms 
of an effective potential. This can be expanded around a nontrivial 
fixed point using the power counting appropriate to a system with large 
scattering length. The terms in the boson self-energy are then in
one-to-one correspondence with the terms in the effective-range expansion.

\section*{Acknowledgements}

I am very grateful to Koji Harada for the invitation to visit Kyushu 
University and to the group there for their hospitality. I also wish 
to thank the following for useful discussions: K. Harada, B. Krippa, 
H. Kubo, D. Litim, J. McGovern, J. Pawlowski and N. Walet.

\end{document}